\documentclass[11pt]{article}
\usepackage{graphicx}
\usepackage{amsmath}

\newcommand{\BABARPubYear}    {08}

\newcommand{\BABARConfNumber} {003}
\newcommand{\SLACPubNumber} {13330}

\newcommand\etal{{\it et al.}}

\newcommand{\UfourS}{\ensuremath{\Upsilon(4S)}}
\def\Bz      {\ensuremath{B^0}\xspace}
\def\Bbar    {\kern 0.18em\overline{\kern -0.18em B}{}\xspace}
\def\BB{\ensuremath{B\Bbar}\xspace} 
\def\Dbar    {\kern 0.18em\overline{\kern -0.18em D}{}\xspace}

\newcommand{\pvec}{{\bf p}}

\newcommand{\mes}{\ensuremath{m_{\rm ES}}}
\newcommand{\de}{\ensuremath{\Delta E}}
\newcommand{\xf}{\ensuremath{{\cal F}}}
\newcommand{\costhr}{\ensuremath{\cos\thetaT}}
\newcommand{\hel}{\ensuremath{{\cal H}}}

\newcommand{\thetaT}{\ensuremath{\theta_{\rm T}}}

\newcommand{\fKuPi}{\ensuremath{K_1\pi}}

\newcommand{\fKunoPi}{\ensuremath{K_1(1270)\pi}}
\newcommand{\fKunopPi}{\ensuremath{K_1(1400)\pi}}
\newcommand{\fKunocPi}{\ensuremath{K_1(1270)^+\pi^-}}
\newcommand{\fKunopcPi}{\ensuremath{K_1(1400)^+\pi^-}}
\newcommand{\fKucPi}{\ensuremath{K_1^+\pi^-}}

\newcommand{\Kuno}{\ensuremath{K_1(1270)}}
\newcommand{\Kunop}{\ensuremath{K_1(1400)}}
\newcommand{\Ku}{\ensuremath{K_1}}

\newcommand{\Kuc}{\ensuremath{K_1^+}}

\newcommand{\BtosumKunocPi}{\ensuremath{\Bz\ra\fKucPi}}
\newcommand{\BtosumKucPi}{\ensuremath{\Bz\ra\fKucPi}}
\newcommand{\rsumKucPi}{\ensuremath{31.0 \pm 2.7 \pm 6.9}}
\newcommand{\BrsumKucPi}{\ensuremath{ (\rsumKucPi) \times 10^{-6} }}

\input pubboard/babarsym
 
\long\def\inst#1{\par\nobreak\kern 4pt\nobreak
    {\it #1}\par\vskip 10pt plus 3pt minus 3pt}

\begin{document}
{\pagestyle{empty}

\begin{flushleft}

\end{flushleft}

\begin{flushright}
\babar-CONF-\BABARPubYear/\BABARConfNumber \\
SLAC-PUB-\SLACPubNumber \\
July 2008 \\
\end{flushright}

\par\vskip 5cm

\begin{center}
\Large  \bf\boldmath Measurement of branching fractions of $\Bz$ decays to
 \fKunocPi and \fKunopcPi
\end{center}
\bigskip

\begin{center}
\large The \babar\ Collaboration\\
\mbox{ }\\
\today
\end{center}
\bigskip \bigskip

\begin{center}
\large \bf Abstract
\end{center}
We present a measurement of the branching fraction of neutral $B$ meson
decaying to
final states containing
a $K_1$ meson, 
i.e. \Kuno\ and \Kunop, 
and a charged pion.
The data, collected with the \babar\ detector at the Stanford Linear
Accelerator Center, represent 454 million \BB\ pairs produced in
\epem\ annihilation. 
We measure the branching fraction  
${\cal{B}}(\BtosumKunocPi) = \BrsumKucPi$, where 
the first error quoted is statistical and  the second is systematic.
In the framework of the $K-$matrix formalism used to describe these 
decays, we also set limits on the ratio of the production constants
for the $K_1(1270)^+$ and $K_1(1400)^+$ mesons in $B^0$ decays.

\vfill
\begin{center}

Submitted to the 34$^{\rm th}$ International Conference on High-Energy Physics, ICHEP 08,\\
30 July---5 August 2008, Philadelphia, Pennsylvania.

\end{center}

\vspace{1.0cm}
\begin{center}
{\em Stanford Linear Accelerator Center, Stanford University, 
Stanford, CA 94309} \\ \vspace{0.1cm}\hrule\vspace{0.1cm}
Work supported in part by Department of Energy contract DE-AC02-76SF00515.
\end{center}

\newpage
} 

%
%
\begin{center}
\small

The \babar\ Collaboration,
\bigskip

%
B.~Aubert,
M.~Bona,
Y.~Karyotakis,
J.~P.~Lees,
V.~Poireau,
E.~Prencipe,
X.~Prudent,
V.~Tisserand
\inst{Laboratoire de Physique des Particules, IN2P3/CNRS et Universit\'e de Savoie, F-74941 Annecy-Le-Vieux, France }
J.~Garra~Tico,
E.~Grauges
\inst{Universitat de Barcelona, Facultat de Fisica, Departament ECM, E-08028 Barcelona, Spain }
% ...

\end{center}\newpage

\section{INTRODUCTION}
\label{sec:Introduction}
$B$ meson decays to final states containing  
an axial-vector meson and a pseudoscalar meson 
have been recently studied both experimentally and
theoretically. Branching fractions of the $B$ mesons decays to final
states containing an $a_1(1260)$ or $b_1$ meson associated with a   
pion or a kaon have been measured experimentally \cite{APDECAYS}. 
Theoretical predictions for the branching fractions of $B$
decay modes to final states containing an axial-vector and a
pseudoscalar meson have been calculated assuming a naive 
factorization hypothesis \cite{LAPORTA,CALDERON} and QCD
factorization~\cite{CHENG}. Expected branching fractions of these $B$
meson decay modes  are of the order of $10^{-6}$.

Recently the \babar\ Collaboration has measured \CP-violating
asymmetries in $B^0 \ra a_1(1260)^{\pm} \pi^{\mp}$ decays and
determined the angle $\alpha_{\rm eff}$~\cite{ALPHA}. 
In the absence of penguin contributions in these decay modes, this
angle would coincide with the angle $\alpha$ of the unitary triangle
of the Cabibbo-Kobayashi-Maskawa 
quark-mixing matrix~\cite{CKM}. 
Theoretical bounds on $\Delta \alpha = \alpha -\alpha_{\rm eff}$ in
these decay modes based on SU(3) flavor-symmetry have been derived
in~\cite{BOUNDS}.
The rates of $B \rightarrow \fKunoPi$ and  
$B \rightarrow \fKunopPi$ 
decays are experimental inputs to the calculation of these bounds. 
For the $\fKunopcPi$ decay mode~\footnote{Except as noted 
explicitly, we use a particle name to denote either 
member of a charge conjugate pair.} 
there exists a published experimental
upper limit at 90\%\ confidence level (CL) of $1.1\times10^{-3}$~\cite{ARGUS}. 
Preliminary results for the branching fractions of the $\fKunocPi$ and 
$\fKunopcPi$ decay modes were obtained by the \babar\ Collaboration
on a sample of 384 million $\BB$ pairs~\cite{MORIOND}.
In the following, we use \Ku\ to indicate both \Kuno\ and \Kunop\ mesons.

\section{THE \babar\ DETECTOR AND DATASET}
\label{sec:babar}

The results presented here are based on a sample of $N_{\BB} = 454.3 \pm 5.0$ 
million $\BB$ pairs collected with the \babar\
detector~\cite{BABARNIM} at the PEP-II $e^+e^-$ asymmetric-energy
storage rings. The $e^+e^-$ center-of-mass energy  $\sqrt{s}$ is equal
to $10.58 \gev$, corresponding to the $\UfourS$ resonance.

Momenta of charged particles are measured in a
tracking system consisting of a silicon vertex tracker with five
double-sided layers and a 40-layer  drift chamber, both within the 1.5
T magnetic field of a solenoid.  
Identification of charged hadrons is provided by measurements
of the energy loss in the tracking devices and by a ring-imaging
Cherenkov detector. 
For lepton identification, we use the energy deposit in a
CsI(Tl) electromagnetic calorimeter and the pattern of hits 
in resistive plate chambers (partially upgraded to limited
streamer tubes for a subset of the data used in this analysis)
intervalled with the passive material comprising the solenoid
magnetic flux return.

\section{ANALYSIS METHOD}
\label{sec:Analysis}
The $\Bz \rightarrow \fKucPi$ candidates are 
identified from the
$\Kuc \rightarrow K^+\pi^+\pi^-$ final state, with reconstructed mass 
$m_{K\pi\pi}$ in the $\left[1.1,1.8\right]$ \gevcc range.
They are kinematically characterized by 
$\mes=[(s/2+\pvec_{\Upsilon}\cdot\pvec_B)^2/E_{\Upsilon}^2-\pvec_B^2]^{1/2}$
and $\de = (E_{\Upsilon}E_B-\pvec_{\Upsilon}\cdot\pvec_B-s/2)/\sqrt{s}$,
where $(E_B,\pvec_B)$ is the four-momentum of the $B$ candidate, and 
$(E_{\Upsilon},\pvec_{\Upsilon})$ is the $e^+e^-$ initial state
four-momentum, both in the laboratory frame.
We require $\mes > 5.25$ \gevcc\ and $|\de| < 0.15$ \gev. 

To reject the dominant $\epem \ra $ quark-antiquark background, we use
the thrust angle \thetaT\ between the $B$-candidate thrust axis and that of
the rest of the event,  calculated in the center-of-mass (CM) frame, 
and a Fisher discriminant \xf~\cite{FISHER}. 
The discriminant combines the polar angles of the $B$-momentum
vector and the $B$-candidate thrust axis with respect to the beam
axis, and 
the zeroth and second moments
of the energy flow around the $B$-candidate
thrust axis, calculated in the CM frame~\cite{FISHER}.

The resonant $K^+\pi^+\pi^-$ system can receive contributions from
several strange resonances in the selected range 
for $m_{K\pi\pi}$, besides $K_1$ mesons, such as $K_1^{*}(1410)^+$ ($J^P=1^-$), 
$K_1^{*}(1680)^+$ ($1^-$), and $K_2^{*}(1430)^+$ ($2^+$). Decays containing
any of these resonances are characterized by different angular distributions.
We define \hel\ as the cosine of the angle between the direction of the
primary pion from $B$ decay and the normal to the plane defined by
\Ku\ daughters in \Ku\ rest frame. We require $|\hel| < 0.95$ to 
reduce background from $B^0 \rightarrow V^+ \pi^-$ decay modes, where $V^+$ is
a vector meson decaying to $K^+\pi^+\pi^-$.

Background from $B$ decays to final states with charm is suppressed by 
rejecting a signal candidate if it has at least one track in common 
with a background $B$ candidate, reconstructed in any of the 
$B^0 \ra D^- \pi^+$, $B^0 \ra D^{*-} \pi^+$, and $B^+ \ra \bar D^0 \pi^+$ 
background decay channels, with $D$ meson mass within $0.07$~\gevcc\ of the 
nominal value (if more than one such background candidates are reconstructed 
per event per background channel, the one with the highest $B$ vertex fit 
$\chi^2$ probability is chosen).  
To suppress background from $B$ decays to final states with
charmonium we calculate the invariant mass of the neutral 
$\pi \pi$ combination of the primary pion from $B$ decay
with the opposite charge pion from \Ku\ decay, and require that it 
is not consistent 
with any of the \ccbar\ mesons $J/\psi$, $\psi(2S)$, $\eta_c$,
$\eta_c(2S)$, $\chi_{c0}(1P)$, and  $\chi_{c1}(1P)$.
We also make particle identification requirements to identify pions and kaons, 
and veto muons, electrons and protons.

The average number of candidates found 
per selected event 
in the data sample is $1.20$.
In case of events with multiple candidates, we select the candidate
with the highest $B$ vertex fit $\chi^2$ probability.
We classify the events according to the invariant masses
of the $K^+ \pi^-$ and $\pi^+ \pi^-$ systems in the $\Kuc \ra K^+ \pi^+
\pi^-$ final state: events which satisfy the requirement 
$0.846 < m_{K\pi} < 0.946$ \gevcc\ belong to class 1 ("$K^*$ region"); 
events not included in class 1 for which $0.500 < m_{\pi\pi} < 0.800$ 
\gevcc\ belong to class 2 ("$\rho$ region"); all other events are rejected.

A two-resonances, six channels $K$-matrix model \cite{KMATRIX} is used 
to describe the resonant $K\pi\pi$ system for the signal~\cite{WA3}.
The notation is consistent with that used in \cite{WA3}. 
The labels $a$ and $b$ in the following paragraphs refer to $K_1(1400)$ 
and $K_1(1270)$, respectively. The production amplitude for channel 
$i = \{(K^*~\pi)_{S-wave}, (K^*~\pi)_{D-wave},$ $\rho~K, K_0^*~\pi, f_0(1370)~K, \omega~K\}$ 
is given by
\begin{equation}
F_i = e^{\mathrm{i} \delta_i}\sum_j(\mathbf{1}-\mathrm{i}\mathbf{K}{\boldsymbol \rho})_{ij}^{-1}
\mathbf{P}_j,
\end{equation}
where
\begin{equation}
K_{ij} = \frac{f_{ai}f_{aj}}{M_a-M}+\frac{f_{bi}f_{bj}}{M_b-M},
\end{equation}
$\delta_i$ are offset phases ($\delta_{(K^*\pi)_S} \equiv 0$), and $\mathbf{P}$ is
the production vector
\begin{equation}
P_i = \frac{f_{pa}f_{ai}}{M_a-M}+\frac{f_{pb}f_{bi}}{M_b-M}.
\end{equation}

The decay  constants $f_{ai}$ , $f_{bi}$  and the $K$-matrix poles
$M_{a}$  and $M_{b}$ are real. 
The elements of the diagonal phase space matrix \mbox{\boldmath$\rho$}
for the process $K_1 \rightarrow 3 + 4$, $3 \rightarrow 5 + 6$, 
where $4$, $5$ and $6$ are long-lived pseudoscalar particles and $3$ is a
resonance,
have been approximated with the form 
\begin{equation}
\rho_{i}(M)=\frac{\sqrt{8}}{M}\left[\frac{m^*m_4}{m^*+m_4}(M-m^*-m_4+\mathrm{i}\Delta)\right]^{1/2},
\end{equation}
where $M$ is the mass of $K_1$,  $m_4$ is the mass of 4, 
$m^*$ is the mean mass of 3 and $\Delta$ is the half width of
$3$ \cite{PHSP}. 
The parameters of $\mathbf{K}$ and the offset phases $\delta_i$ are obtained
from a fit to
the intensity and the relative phases of the $K\pi\pi$ channels,
which were extracted by the ACCMOR Collaboration in a partial wave analysis of 
the data on the reaction $K^-p \rightarrow K^-\pi^+\pi^-p$ 
accumulated by the WA3 experiment~\cite{WA3}.
For the fit to WA3 data we add a background term to the production 
vector~\cite{DECK}. 
The decay constants for the $\omega~K$ channel are fixed according to the 
quark model \cite{WA3}. 

We express the complex production constants $f_{pa}$ and $f_{pb}$ in terms
of the production parameters ${{\boldsymbol \zeta}=(\theta,\phi)}$:
$f_{pa}\equiv \cos\theta$, $f_{pb}\equiv \sin\theta e^{\mathrm{i}\phi}$, where
$\theta \in [0,\pi/2]$, and $\phi \in [0,2\pi]$. In this parameterization,
$\tan\theta$ represents the magnitude of the production constant for 
the $K_1(1270)$ meson relative to that for the $K_1(1400)$ meson, while
$\phi$ is the relative phase.

Signal MC samples are generated by weighting the $(K^+\pi^+\pi^-)\pi^-$
population according to the amplitude
$\sum_{i\neq \omega K}\langle K^+\pi^+\pi^-| i \rangle F_i$, where the term 
$\langle K^+\pi^+\pi^-| i \rangle$
consists of a factor describing the angular distribution of the 
$(K^+\pi^+\pi^-)$ system resulting from \Ku\ decay, 
an amplitude for the resonant $\pi^+\pi^-$ and $K^+\pi^-$ systems, 
and isospin factors, and is 
calculated using the formalism described in \cite{HERNDON}. 
The branching fraction for $\Ku \ra \omega K$ is
accounted for as a correction to the total selection efficiency.

We use an unbinned,  extended maximum-likelihood (ML) fit to extract the
event yields $n_{s,r}$ and the parameters of the probability density function
(PDF) ${\cal P}_{s,r}$. The subscript $r=\{1,2\}$ corresponds to one of the
event classes defined above. The index $s$ represents six event
categories used in our data model: 
\begin{itemize}
\item the signal $\Bz \rightarrow \fKucPi$ ($s=1$), 
\item possible backgrounds from ${B^0\ra  a_1(1260)^{\pm} \pi^{\mp} 
\rightarrow (\pi^{\pm} \pi^+ \pi^-) \pi^{\mp}}$ ($s=2$),  
\item $B^0 \ra D^- \pi^+ \rightarrow (K^+ \pi^- \pi^-) \pi^{+}$ ($s=3$), 
\item $B^0 \ra K^{*}(1410)^+ \pi^-$ ($s=4$), 
\item $B^0 \ra K^{*0}\pi^+\pi^- + \rho^0 K^+ \pi^-$ ($s=5$),
\item combinatorial background ($s=6$). 
\end{itemize}
We perform a likelihood scan with respect to the parameters 
${\boldsymbol \zeta}$, with $21\times 21$ points. At each point, a 
simultaneous fit to the two event classes is performed.

The signal and background PDFs are the products of the PDFs for 
independent variables. 
The signal PDFs for $\de$, $\mes$, and $\xf$ are parameterized as 
the sum of Gaussian functions for the core of the distributions
plus empirical functions accounting for the tails.
The dependence on {\boldmath$\zeta$} of the selection efficiencies 
and the signal PDF for $m_{K\pi\pi}$ are parameterized by means of
templates modeled upon signal MC samples. 
Resonance production occurs in the non-signal $B$ background and is taken 
into account in the PDFs.
For the combinatorial background, we use polynomials, except for
$\mes$ and $\xf$ distributions which are parameterized by
an empirical phase-space function \cite{ARGUSMES} and by Gaussian functions,
respectively. 
The combinatorial background PDF is found to describe well both the
dominant quark-antiquark background and the background from random 
combinations of $B$ tracks.
For all components, PDFs for ${\cal H}$ are parameterized with polynomials.

The likelihood ${\cal L}_e$ for each candidate $e$ 
belonging to class $r$ is defined as
${\cal L}_e = \sum_{s}n_{s,r}\, {\cal P}_{s,r}$({\boldmath ${\rm
    x}_e$};~{\boldmath$\zeta$},~{\boldmath$\xi$}),
where the PDFs are formed using the set of observables
{\boldmath ${\rm x}_e$}~$=\{\de$, $\mes$, $\xf$, $m_{K\pi\pi}$, $\hel$\}
and the dependence on production parameters {\boldmath$\zeta$} is
relevant only for the signal PDF. 
{\boldmath$\xi$} represents all other PDF parameters. 
In the definition of ${\cal L}_e$ the yields of the signal category
for the two classes are expressed as a function of the signal branching
fraction $ {\cal B}$ as
$n_{1,1}={\cal B} \times N_{\BB} \times
\epsilon_1$({\boldmath$\zeta$}) and $n_{1,2}={\cal B} \times N_{\BB}
\times \epsilon_2$({\boldmath$\zeta$}), where  the total selection
efficiency, $\epsilon_r$({\boldmath$\zeta$}), includes the daughter
branching fractions and the reconstruction efficiency obtained from MC
simulation. 

The signal branching fraction is a free parameter in the fit.
The yields for event categories $s=2$ and $3$ are fixed to
the values estimated from MC. The yields for the other background 
components are determined from the fit. 
The PDF parameters for combinatorial background are left free to vary in 
the fit while those for the other event categories
are fixed to the values extracted from  
Monte Carlo (MC) simulation~\cite{GEANT} and calibration
$B^0 \ra D^-\pi^+$ decays.

\section{SYSTEMATIC STUDIES}
\label{sec:Systematics}
The main sources of systematic uncertainties are summarized in
Table~\ref{tab:systtab}. 
We repeat the fit by varying all the parameters in {\boldmath$\xi$} 
which were not left floating in the fit within their uncertainties,
and obtain the associated systematic uncertainties. 
The signal PDF model excludes the fake combinations originating from
mis-reconstructed signal events.
The biases due to the presence of fake combinations, 
or other imperfections in the signal PDF model are estimated with MC
simulation.
The finite resolution of the likelihood scan is also a source of
bias. 
A systematic error is evaluated by varying the $K_1(1270)$ and $K_1(1400)$ 
mass poles in the signal model, the parameterization of the intermediate 
resonances in $K_1$ decay, and the offset phases $\delta_i$.
Additional systematic uncertainty originates from potential peaking 
\BB\ background, including $B^0 \ra K_2^{*}(1430)^+ \pi^-$ and  
$B^0 \ra K_1^{*}(1680)^+ \pi^-$, and is evaluated by introducing
the corresponding components 
in the definition of the likelihood 
and repeating the fit with their yields fixed to 
values estimated from the available experimental information~\cite{PDG}.
We assign a systematic uncertainty due to yield variation
in the $B^0\ra  a_1(1260)^{\pm} \pi^{\mp}$ and  $B^0 \ra D_{K^+ \pi^- \pi^-}^-
\pi^+$ event categories. 
The above systematic uncertainties do not scale with event yield
and are included in the calculation of the significance of the result.

We estimate the systematic uncertainty due to the interference between 
the $B^0 \rightarrow \fKucPi$ and the $B^0 \ra K^{*0}\pi^+\pi^- + \rho^0 
K^+ \pi^-$ decays using simulated samples in which the decay amplitudes 
are generated according to the results of this measurement. The overall 
phases and relative contribution for $K^{*0}\pi^+\pi^-$ and $\rho^0 K^+ \pi^-$
interfering states are assumed to be constant across the phase space
and varied between zero and a maximum value using uniform prior 
distributions. We calculate the systematic uncertainty from the RMS 
variation of the average signal branching fraction and parameters.
In the calculation of significance, this effect is assumed to scale 
with the square root of the signal branching fraction.
The systematic uncertainties in efficiencies are dominated 
by those in track finding and particle identification.
Other systematic effects arise from event-selection criteria,
such as track multiplicity and thrust angle,
and the number of $B$ mesons.        
\begin{table}[h]   
\begin{center}
\caption{Estimates of systematic errors. 
For the branching fraction, some of these errors are additive (A) 
and given in units of $10^{-6}$, others are
multiplicative (M) and given in \% . Contributions are combined in
quadrature.} 
\label{tab:systtab}
\begin{tabular}{lccc}
\hline\hline
Quantity                       & ${\cal B}$ & $\theta$ & $\phi$   \\

\hline
PDF parameters (A)                & $ 1.0 $ & $ 0.01 $ & $ 0.04 $ \\
MC/data correction (A)            & $ 1.2 $ & $ 0.05 $ & $ 0.27 $ \\   
ML Fit bias   (A)                 & $ 0.6 $ & $ 0.03 $ & $ 0.02 $ \\  
Scan   (A)                        & $ 1.3 $ & $ 0.04 $ & $ 0.16 $ \\  
$K_1$ mass poles (A)              & $ 2.2 $ & $ 0.01 $ & $ 0.36 $ \\ 
$K_1$ offset phases (A)           & $ 0.2 $ & $ 0.01 $ & $ 0.02 $ \\ 
$K_1$ intermediate resonances (A) & $ 0.5 $ & $ 0.00 $ & $ 0.06 $ \\ 
Peaking $\BB$ bkg (A)             & $ 0.8 $ & $ 0.02 $ & $ 0.27 $ \\
Fixed background yields (A)       & $ 0.8 $ & $ 0.04 $ & $ 0.08 $ \\ 
Interference (A)                  & $ 5.9 $ & $ 0.25 $ & $ 0.52 $ \\ 
MC statistics (M)                 & $ 1.0 $ & $ -    $ & $ -    $ \\
Particle ID (M)                   & $ 2.9 $ & $ -    $ & $ -    $ \\
Track finding (M)                 & $ 1.0 $ & $ -    $ & $ -    $ \\
\costhr       (M)                 & $ 1.0 $ & $ -    $ & $ -    $ \\
Track multip.  (M)                & $ 1.0 $ & $ -    $ & $ -    $ \\
Number \BB\ (M)                   & $ 1.1 $ & $ -    $ & $ -    $ \\ 
\hline
Total (A)                         & $ 6.9 $ & $ 0.26 $ & $ 0.76 $ \\
\hline\hline
\end{tabular}
\end{center}
\end{table}

\section{RESULTS}
\label{sec:Results}
\begin{figure}[!htb]
\begin{center}
  \includegraphics[width=0.445\linewidth]{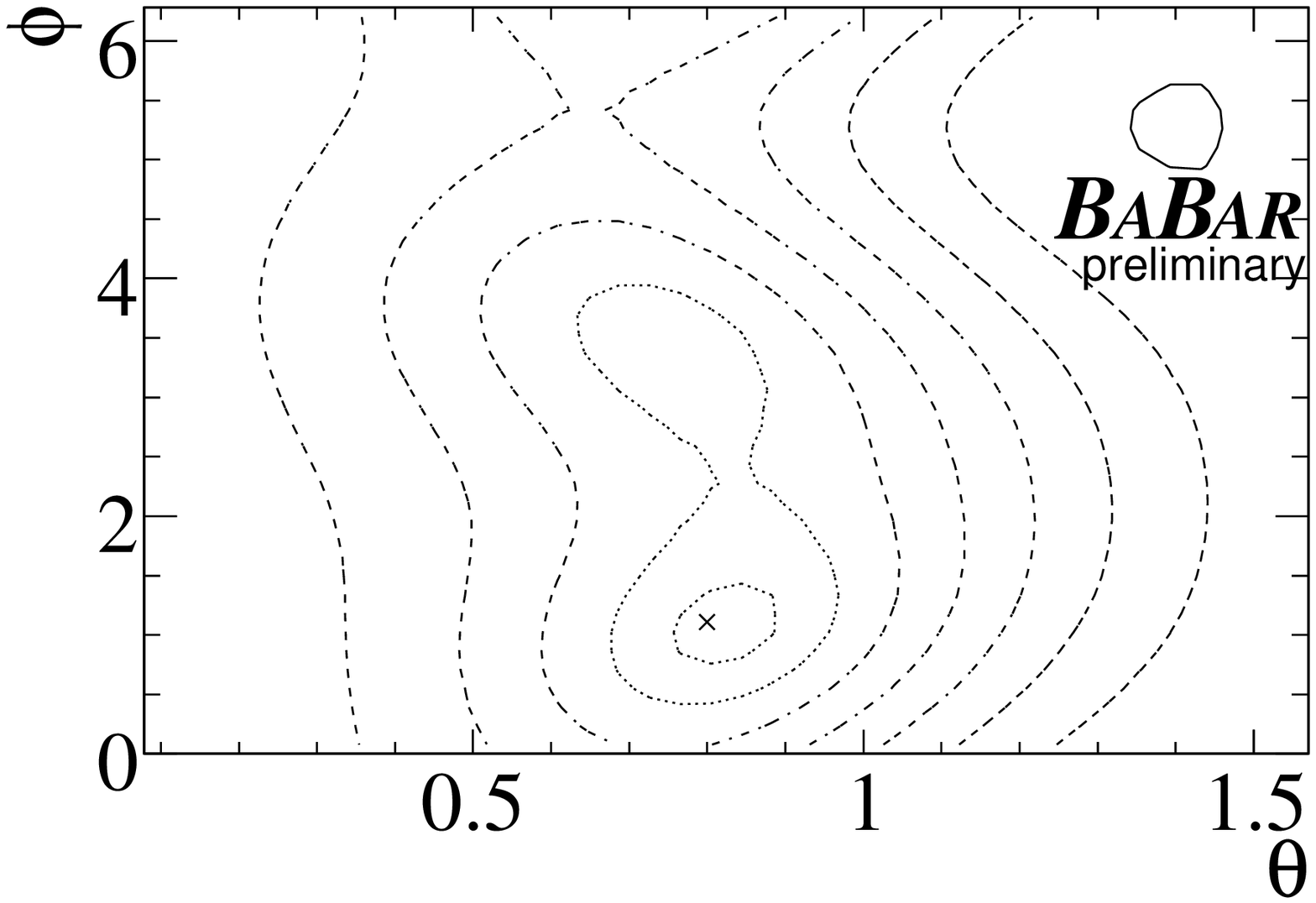}
  \includegraphics[width=0.445\linewidth]{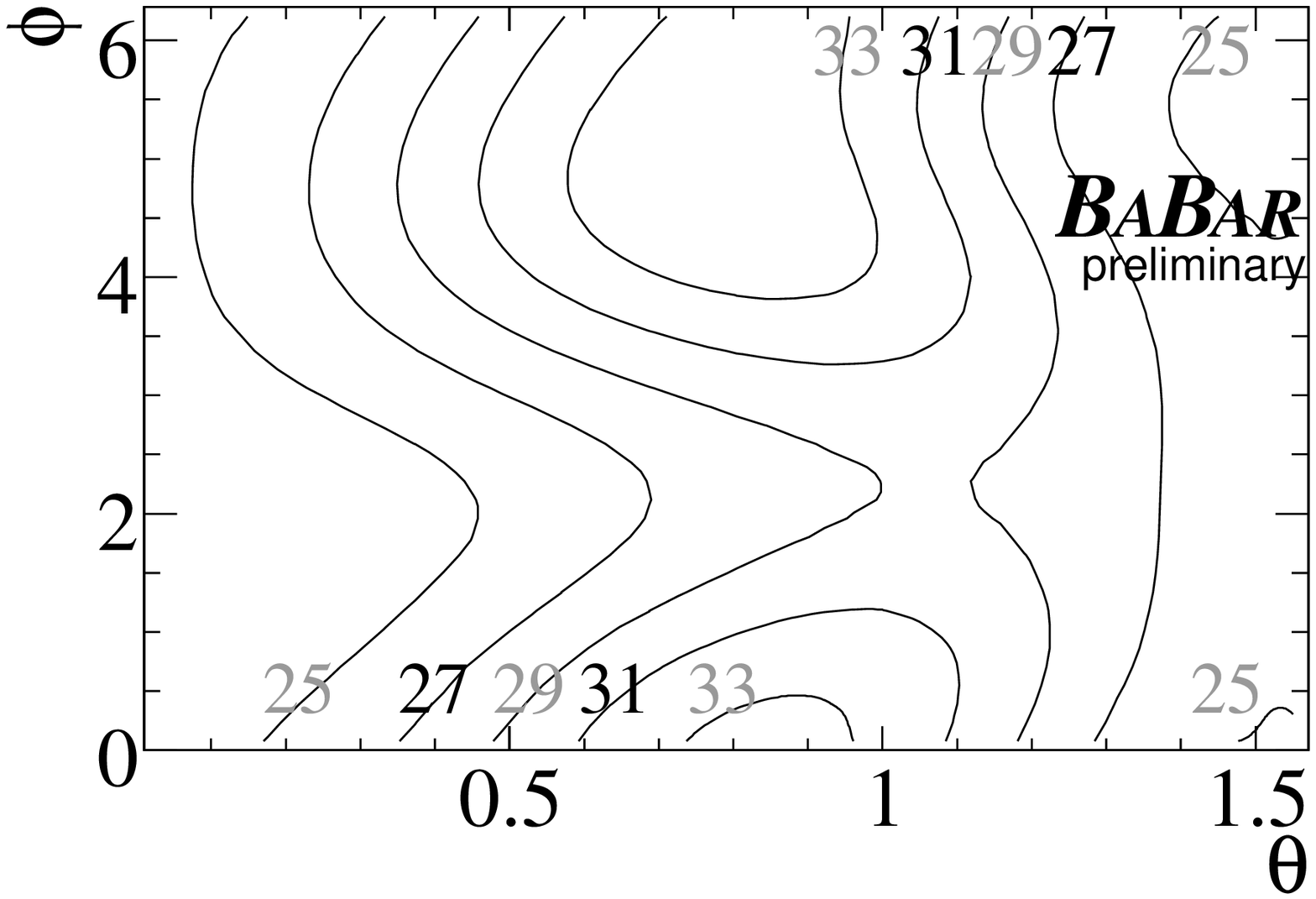}
  \caption{Left: contour plot of $-\ln{\cal L}$ (no systematic effects 
    included) in the $\theta,\phi$ plane; each line corresponds to a $n^2/2$ 
    increase in $-\ln{\cal L}$, with $n=\{1,2,...\}$, with respect to the 
    minimum (indicated by a cross). Right: fitted value of ${\cal B}_{sg}$, in
    units of $10^{-6}$ as a function of $\theta$ and $\phi$.}
  \label{fig:nllscan}
  \end{center}
\end{figure}

Figure~\ref{fig:nllscan} shows the likelihood scan and the values
of ${\cal B}_{sg}$ which minimize $-\ln{\cal L}$ as a function of $\theta$ 
and $\phi$.
The absolute minimum
occurs at $\theta = 0.785$ and $\phi = 0.942$, and the signal branching
fraction corresponding to that point of the scan is 
${\cal{B}}(\BtosumKunocPi) = ( 31.0 \pm 2.7 ) \times 10^{-6}$. 
By interpolation between neighbouring
points of the likelihood scan we extract $\theta = 0.81 \pm 0.06$
and $\phi = 1.11 \pm 0.28$. The quoted errors on the branching fraction 
and production parameters ${\bf \zeta}$ are only statistical and correspond
to a $0.5$ increase in $-\ln{\cal L}$. A second, local minimum is located 
at $\theta = 0.785$ and $\phi = 3.454$, and is associated to a $1.0$ 
increase in $-\ln{\cal L}$.

A conservative estimate of significance is calculated from a likelihood ratio 
test $\Delta(-2\ln{\cal L}) $, assuming a $\chi^2$ distribution with $N=3$ 
degrees of freedom and minimizing the significance with respect to the 
production parameters $(\theta,\phi)$. Here $ \Delta(-2\ln{\cal L}) $ is the 
difference between the value of $-2\ln{\cal L}$ 
for zero signal and the value at its minimum for
given values of ${\boldsymbol \zeta}$
(${\cal L}$ represents the convolution of the likelihood with a 
Gaussian function representing additive systematic uncertainties on the 
branching fraction).
We observe a non zero $B^0 \rightarrow \fKucPi$ branching fraction with
significance greater than $5.1~\sigma$.

\begin{figure}[t]
  \begin{center}
  \includegraphics[width=0.700\linewidth]{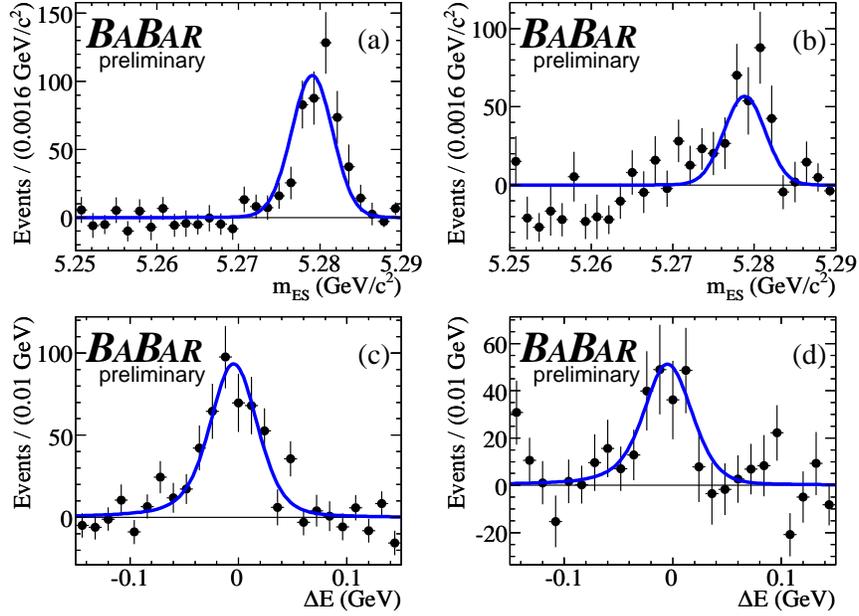}
 \caption{sPlot projections onto a) \mes\ (class 1),
   b) \mes\ (class 2), c) \de\ (class 1),
   d) \de\ (class 2) in the \fKuPi\ decay.
   Points represent on-resonance data, solid line is the signal fit
   function.
 } 
 \label{fig:splots}
  \end{center}
\end{figure}
\begin{figure}[!t]
  \begin{center}
  \includegraphics[width=0.700\linewidth]{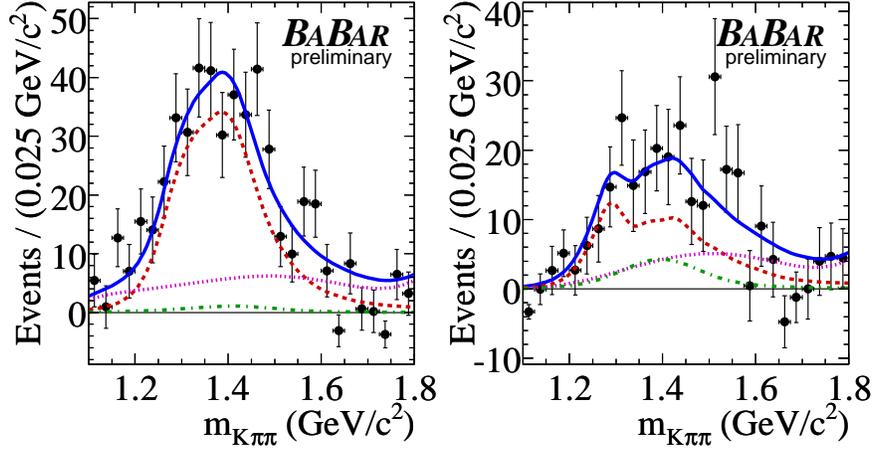}
  \caption{sPlot projection onto $m_{K \pi \pi}$ for class 1 (left) and
    class 2 events (right). 
    Points represent
    on-resonance data, solid line is the sum of the fit functions of the
    decay modes $K_1(1270) \pi + K_1(1400) \pi$ (dashed), 
    $K^*(1410) \pi$ (dash-dotted), and $K^*(892) \pi \pi$ (dotted).
    Here the points are obtained without using any information about
    resonances in the fit, \emph{i.e.} we use only \mes, \de, and \xf\
    variables, while for the normalization of the curves we use the signal 
    yields obtained from the nominal fit. 
  }
  \label{fig:mass}
  \end{center}
\end{figure}
Figure~\ref{fig:splots} shows the distributions of \de\  and  \mes\
for the signal events, obtained by the 
event-weighting technique (sPlot) described in~\cite{SPLOT}. 
For each event, a weight to be signal or background is derived according
to the results of the fit to all variables and the probability distributions 
in the restricted set of variables, in which the projection variable is
omitted. Using these weights, the data is then plotted in the projection 
variable. We show in Figure~\ref{fig:mass} the projection onto 
$m_{K\pi\pi}$.

\begin{figure}[!t]
\begin{center}   
\includegraphics[width=0.445\linewidth]{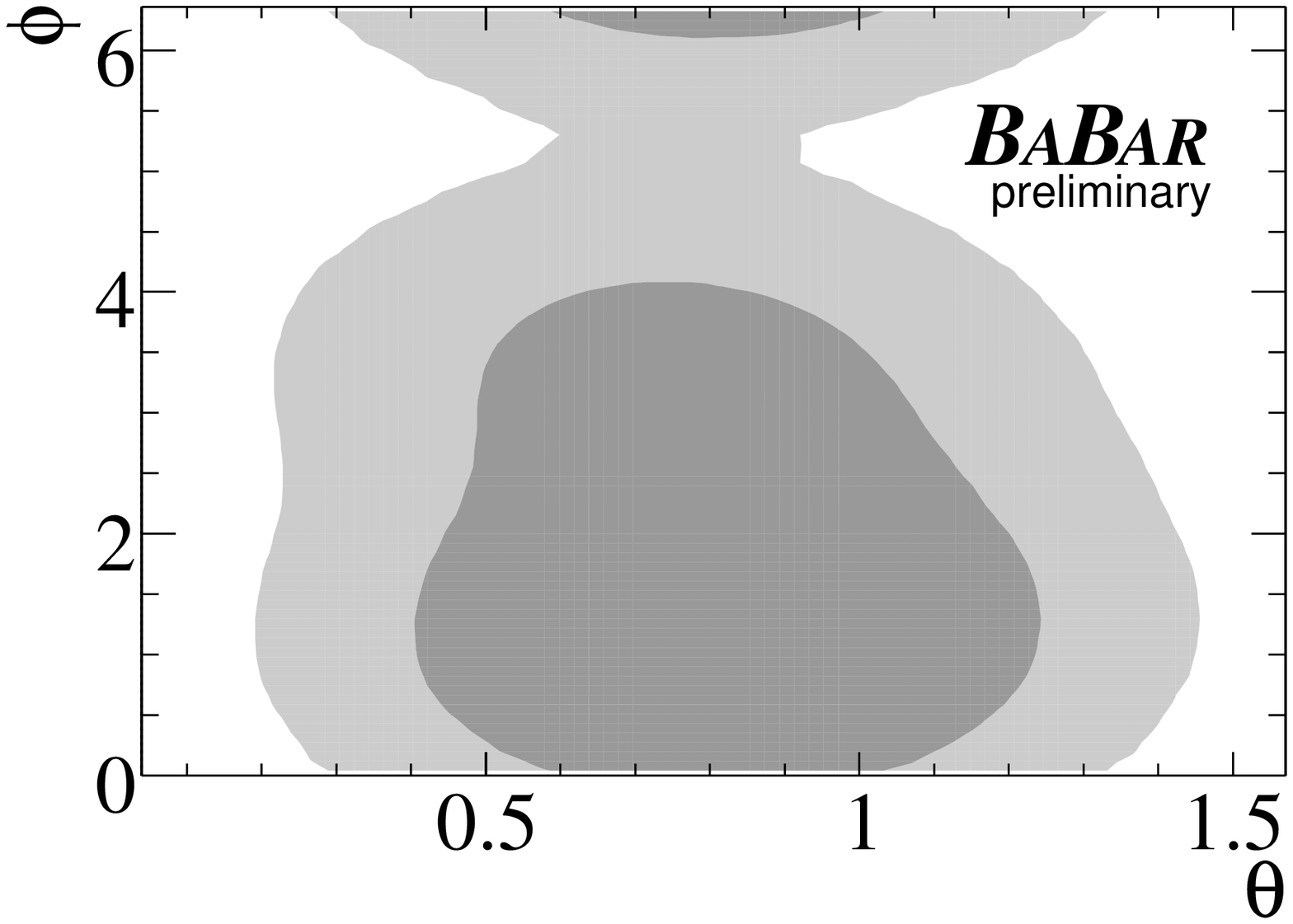}\\
\includegraphics[width=0.445\linewidth]{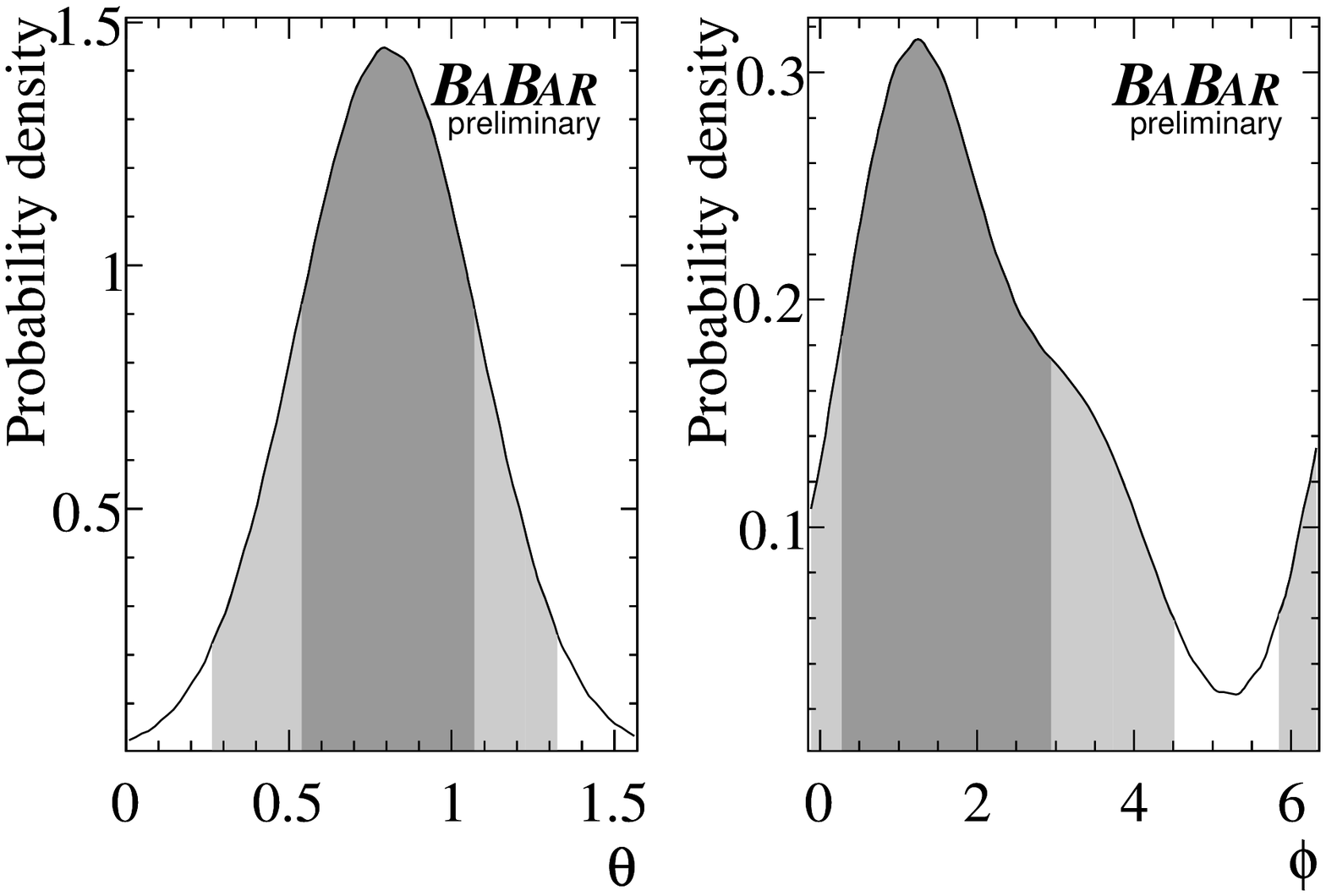}
\caption{68 \% (dark shaded zone) and 95 \% (light shaded zone) probability 
regions for $\theta$ and $\phi$ (top), $\theta$ (bottom-left) and $\phi$ 
(bottom-right).}
\label{fig:clregions}
\end{center}
\end{figure}
The experimental two-dimensional likelihood $\mathcal{L}$ for 
$\theta$ and $\phi$ is convoluted with a two-dimensional Gaussian 
that accounts for the systematic uncertainties.
In Figure~\ref{fig:clregions} we show the distributions we obtain 
for $\theta$, $\phi$ and $\theta$ vs. $\phi$ (the 68\% and 95\% probability 
regions are shown in dark and light shading respectively, and
are defined as the regions which satisfy ${\cal {L}}(r)>{\cal {L}}_{min}$ and 
$\int_{{\cal {L}}(r)>{\cal {L}}_{min}}{\cal {L}}(r) dr = 68\%~~(95\%)$, 
where $r$ is the projected set of variables).
The condition ${\cal {L}}(r)>{\cal {L}}_{II}$,
where ${\cal {L}}_{II}$ is the value of the likelihood evaluated at the 
position of the second, local minimum in Figure~\ref{fig:nllscan}, defines
a $48\%$ probability region, with systematic uncertainties included, on the 
$\theta$ vs. $\phi$ plane.

\section{CONCLUSIONS}
\label{sec:Conclusions}
We measure the branching fraction
\begin{eqnarray}
{\cal{B}}(\BtosumKucPi) = ( \rsumKucPi ) \times 10^{-6}, \nonumber
\end{eqnarray}
with significance greater than $5.1~\sigma$.
The first error quoted is statistical and the second systematic.
The value of the branching fraction measured in this analysis is consistent 
with preliminary results obtained by \babar\ Collaboration~\cite{MORIOND},
and is to be compared with the naive factorization \cite{LAPORTA,CALDERON} 
and QCD factorization~\cite{CHENG} estimates, of order $10^{-6}$.

For the production parameters we obtain
\begin{eqnarray}
0.25  & < \theta < & 1.32 \nonumber \\
-0.51 & < \phi   < & 4.51 \nonumber
\end{eqnarray}
at $95\%$ probability. 
This analysis represents the first attempt to measure
the relative phase between the production amplitudes of 
\Kuno\ and \Kunop\ mesons in $B$ decays.

\section{ACKNOWLEDGMENTS}
\label{sec:Acknowledgments}

We are grateful for the 
extraordinary contributions of our \pep2\ colleagues in
achieving the excellent luminosity and machine conditions
that have made this work possible.
The success of this project also relies critically on the 
expertise and dedication of the computing organizations that 
support \babar.
The collaborating institutions wish to thank 
SLAC for its support and the kind hospitality extended to them. 
This work is supported by the
US Department of Energy
and National Science Foundation, the
Natural Sciences and Engineering Research Council (Canada),
the Commissariat \`a l'Energie Atomique and
Institut National de Physique Nucl\'eaire et de Physique des Particules
(France), the
Bundesministerium f\"ur Bildung und Forschung and
Deutsche Forschungsgemeinschaft
(Germany), the
Istituto Nazionale di Fisica Nucleare (Italy),
the Foundation for Fundamental Research on Matter (The Netherlands),
the Research Council of Norway, the
Ministry of Education and Science of the Russian Federation, 
Ministerio de Educaci\'on y Ciencia (Spain), and the
Science and Technology Facilities Council (United Kingdom).
Individuals have received support from 
the Marie-Curie IEF program (European Union) and
the A. P. Sloan Foundation.

\end{document}